\documentclass{article}
\usepackage{graphicx}
\usepackage[round]{natbib}
\usepackage{epsfig}
\title{
Year ahead prediction of US landfalling hurricane numbers: the
optimal combination of multiple levels of activity since 1900
 }

\author{
Roman Binter (LSE)\\
Stephen Jewson (RMS)\footnote{\emph{Correspondence email}: \texttt{stephen.jewson@rms.com}}\\
Shree Khare (RMS)\\
Adam O'Shay (RMS)\\
Jeremy Penzer (LSE)\\}

\begin{document}
\maketitle
\begin{abstract}
In earlier work we considered methods for predicting future levels
of hurricane activity based on the assumption that historical mean
activity was at one constant level from 1900 to 1994, and has been
at another constant level since then. We now make this model a
little more subtle, and account for the possibility of four
different levels of mean hurricane activity since 1900.
\end{abstract}

\section{Introduction}

We want to predict the number of hurricanes that might occur
during the next hurricane season, and we want to make that
prediction as soon as the current season is over. We call this
problem `year-ahead' prediction of hurricane numbers, to
distinguish it from `seasonal' predictions of hurricane numbers,
which are generally made from a point in time closer to the
hurricane season being predicted. Seasonal predictions often use
predictors in the atmosphere and ocean that correlate with
hurricane activity (see, for example, \citet{saundersl05},
\citet{gray84b}, \citet{gray92}, \citet{gray93} and
\citet{gray94}, \citet{elsners93}, \citet{landsea98},
\citet{landsea00}, \citet{lehmiller}). These predictors are
generally less useful, however, at the long lead time that we are
considering, and we therefore take a rather different
approach, based on the idea of using linear combinations of the
historical hurricane number data as a prediction of the future
number of hurricanes. The particular linear combinations we
consider put different weights on different periods of historical
data according to how similar those data are to the current levels
of activity.


Our first attempt in this direction (described in~\citet{j90}) was to assume that
historical hurricane activity was at one (constant) level of
mean activity from 1900 to 1994, and jumped to another (constant) level of mean activity
from 1995 to 2005. We then assumed that 2006 would remain at the new
level of mean activity, and predicted this
new level of activity using a combination of the estimated levels of
mean activity from 1900 to 1994 and 1995 to 2005. We formulated the
problem mathematically as a mean-variance trade-off: the recent data
is (according to our assumptions) unbiased, but too short to give
accurate estimates. The earlier data is biased, but long enough to
make our estimates much more accurate. An appropriately weighted
combination of the data from the two periods can then give a
prediction with MSE lower than the predictions from either period
individually, and we derived expressions that give an estimate of
the weights needed to make this combination.

Perhaps the biggest objection that one might make to this initial
study (in terms of realism), is that no-one really thinks that
hurricane activity was at one constant level from 1900-1994. A
more common point of view is that activity levels were low in the
early part of the 20th century, higher in the middle of the
century, low again until the mid 1990s, and then high again. The
goal of this article is to extend our analysis to account for this
idea. Instead of modelling hurricane numbers in two periods of
constant levels of activity, we now model it in four periods.
Otherwise, our method is exactly as before: we consider how to
make the combinations of the estimated levels of activity in these
four periods, with the goal of minimising the MSE of predictions
for next year, assuming that next year remains at the same level
of mean activity as the current period. Within this framework we
consider a number of different ways of weighting the four periods
of activity.


The periods we use are:
\begin{itemize}
  \item period 4: 1900-1942
  \item period 3: 1943-1964
  \item period 2: 1965-1994
  \item period 1: 1995-2005
\end{itemize}

The definitions of these periods are taken from the change-point analysis in~\citet{elsnerj00},
figure 2. It would be interesting to investigate how robust these
periods are: however, the point of this article is to demonstrate
how to combine data, given the definitions of the four periods,
rather than to investigate the definitions of the periods
themselves. We number the periods in reverse chronological order to emphasize that
period 1 is unique in that we are assuming that 2006 will be at the
same activity level as period 1.

The rest of this paper is as follows.
In section~\ref{data} we present the hurricane data we will use for our analysis.
In section~\ref{four} we define the four period model and derive an expression for the MSE
score we use.
In section~\ref{derive} we discuss how the weights in the model can be calculated.
We then move on to applying various models to the hurricane data in sections~\ref{results_all},
\ref{results_intense} and~\ref{results_basin}.
Finally in section~\ref{conclusions} we discuss our results.

\section{Data}\label{data}

We now describe the hurricane data we will use for this study. The
basic data set is HURDAT~\citep{hurdat}.
We extract the number of hurricanes making landfall each year, in each
category, counting only the most intense landfall of storms that
make multiple landfalls (this is the SSS variable in HURDAT).
Table~\ref{data1} shows the number of hurricanes in each of our
four periods, for all hurricanes, and for increasingly intense
hurricanes. Table~\ref{data2} shows the same data, but now for
numbers of hurricanes per year.

\begin{table}[h!]
  \centering
\begin{tabular}{|c|c|c|c|c|c|c|}
 \hline
 1 & 2 & 3 & 4 & 5 & 6 & 7\\
 \hline
  &  00-42&   43-64& 65-94 & 95-05 & 00-42+65-94 & 43-64+95-05 \\
 \hline
1-5& 76&  43&   38&   24&   101&   68\\
2-5& 46&  27&   21&   17&   67&   44\\
3-5& 27&  18&   14&   10&   41&   28\\
4-5&  8&   5&    3&    1&   11&   6\\
5&    1&   0&    2&    0&    3&   0\\
 \hline
\end{tabular}
\caption{The columns show:
(1) the categories of hurricanes being considered,
(2) the number of landfalling hurricanes in these categories between 1900 and 1942,
(3) the number of landfalling hurricanes in these categories between 1943 and 1964,
(4) the number of landfalling hurricanes in these categories between 1965 and 1994,
(5) the number of landfalling hurricanes in these categories between 1995 and 2005,
(6) the number of landfalling hurricanes in these categories between 1900 and 1942 plus the number between 1965 and 1994,
(7) the number of landfalling hurricanes in these categories between 1943 and 1964 plus the number between 1995 and 2005.
}\label{data1}
\end{table}

\begin{table}[h!]
  \centering
\begin{tabular}{|c|c|c|c|c|c|c|}
 \hline
 1 & 2 & 3 & 4 & 5 & 6 & 7\\
 \hline
  &  00-42&   43-64& 65-94 & 95-05 & 00-42+65-94 & 43-64+95-05 \\
 \hline
 1-5& 1.77& 1.95&   1.27&   2.27&   1.38&   2.06\\
 2-5& 1.07& 1.23&   0.70&   1.55&   0.92&   1.33\\
 3-5& 0.63& 0.82&   0.47&   0.91&   0.56&   0.85\\
 4-5& 0.19& 0.23&   0.10&   0.09&   0.15&   0.18\\
   5& 0.02& 0.00&   0.07&   0.00&   0.04&   0.00\\
\hline
\end{tabular}
\caption{
This table has the same contents as table~\ref{data1}, but the hurricane numbers
are now given as average numbers per year, rather than total numbers.
}\label{data2}
\end{table}

\section{The four period model}\label{four}

In this section we define the statistical model that
represents hurricane activity as having been driven by four
periods of differing underlying mean activity rates.
In fact, we will consider
the general case of $k$ periods, and specialize to $k=4$ later.

We will be taking an approach in which the
historical data is considered as a single realisation of a random variable.
All expectations are then with respect to this random variable.
The random variable we use is:

\begin{equation}
Y_{i,j}             \sim \mbox{Pois}(\lambda_{i}),  (i=1,2,3,k; j=1,...,n_i)
\end{equation}

$Y_{i,j}$ represents the hurricane counts during the $j'$th year of
the $i'th$ period. As mentioned above, we count our periods
backwards in time, so that $j=1$ corresponds to the most recent
period, $j=2$ the one before that, and so on.

Our predictions will be based on averages of the hurricane numbers during these $k$
periods, which we write as:

\begin{equation}
\hat{\lambda}_{i}   =    \frac{1}{n_{i}}\sum_{j=1}^{n_{i}}Y_{i,j} \\
\end{equation}

We are trying to predict the hurricane count for next year, on the
assumption that the hurricane activity next year is at the same
level as the last few years. We are thus trying to predict the next
value in period 1, which we write as $Y_{1,n_1+1}$.

Before considering weighted combinations of the hurricane number averages $\hat{\lambda}_{i}$
during these four periods, we first consider
the two most obvious predictions for $Y_{1,n_1+1}$, which are the average of the hurricane counts in period 1:
\begin{equation}
\hat{Y}_{1,n_{1}+1} =    \hat{\lambda}_{1}
\end{equation}

(we often call this the `short baseline predictor')

and the average of the hurricane counts over all four periods:
\begin{equation}
Y_{1,n_{1}+1}^{\dag}=    \sum_{i=1}^{k} \frac{n_{i}}{n}\hat{\lambda}_{i}
\end{equation}

where $n=\sum_{i=1}^{j} n_i$

(we often call this the `long baseline predictor').

What are the properties of these two basic predictors?

\subsection{Properties of the basic predictors}

Rewriting the two basic predictors in terms of the real (unknown) hurricane
rates in the $k$ periods, plus (also unknown) noise, gives:

\begin{eqnarray}
\hat{Y}_{1,n_{1}+1}  &=& \lambda_{1}+\frac{1}{n_{1}}\sum_{j=1}^{n_{1}}\varepsilon_{1,j}\\
Y_{1,n_{1}+1}^{\dag} &=&
\frac{n_{1}}{n}\left(\lambda_{1}+\frac{1}{n_{1}}\sum_{j=1}^{n_{1}}\varepsilon_{1,j}\right)
+
\frac{n_{2}}{n}\left(\lambda_{2}+\frac{1}{n_{2}}\sum_{j=1}^{n_{1}}\varepsilon_{2,j}\right)
\\&&+
\ldots
+
\frac{n_{k}}{n}\left(\lambda_{k}+\frac{1}{n_{k}}\sum_{j=1}^{n_{1}}\varepsilon_{k,j}\right)
\end{eqnarray}.

The prediction errors for these basic predictions are then:

\begin{eqnarray}
Y_{1,n_{1}+1}-\hat{Y}_{1,n_{1}+1}
&=&
\lambda_{1}+\varepsilon_{1,n_{1}+1}
-
\left(\lambda_{1}+\frac{1}{n_{1}}\sum_{j=1}^{n_{1}}\varepsilon_{1,j}\right) \\
Y_{1,n_{1}+1}-Y_{1,n_{1}+1}^{\dag}
&=&
\lambda_{1}+\varepsilon_{1,n_{1}+1}
-
\frac{n_{1}}{n}\left(\lambda_{1}+\frac{1}{n_{1}}\sum_{j=1}^{n_{1}}\varepsilon_{1,j}\right)
-
\frac{n_{2}}{n}\left(\lambda_{2}+\frac{1}{n_{2}}\sum_{j=1}^{n_{1}}\varepsilon_{2,j}\right) \\
&&-
\ldots
-
\frac{n_{k}}{n}\left(\lambda_{k}+\frac{1}{n_{k}}\sum_{j=1}^{n_{1}}\varepsilon_{k,j}\right) \\
&=&
\frac{1}{n}[(n-n_{1})\lambda_{1}-n_{2}\lambda_{2}-\ldots-n_{k}\lambda_{k}] \\
&&+
\varepsilon_{1,n_{1}+n}
-
\frac{1}{n}\left(\sum_{j=1}^{n_1}\varepsilon_{1,j}
+
\sum_{j=1}^{n_2}\varepsilon_{2,j}
+
\ldots
+
\sum_{j=1}^{n_k}\varepsilon_{k,j}\right)
\end{eqnarray}.

and the statistical properties of these predictions are:

\begin{eqnarray}
E\left(Y_{1,n_{1}+1}-\hat{Y}_{1,n_{1}+1}\right) &=&0 \\
E\left(E\left(Y_{1,n_{1}+1}\right)-\hat{Y}_{1,n_{1}+1}\right) &=&0 \\
\nonumber\\
E\left(Y_{1,n_{1}+1}-Y_{1,n_{1}+1}^{\dag}\right)
&=&
\frac{1}{n}\left[(n-n_{1})\lambda_{1}-n_{2}\lambda_{2}-\ldots-n_{k}\lambda_{k}\right] \\
E\left(E\left(Y_{1,n_{1}+1}\right)-Y_{1,n_{1}+1}^{\dag}\right)
&=&
\frac{1}{n}\left[(n-n_{1})\lambda_{1}-n_{2}\lambda_{2}-\ldots-n_{k}\lambda_{k}\right]
\end{eqnarray}.

\begin{eqnarray}
\mbox{Var}\left(Y_{1,n_{1}+1}-\hat{Y}_{1,n_{1}+1}\right)
&=&
\left(1+\frac{1}{n_{1}}\right)\lambda_{1}\\
\mbox{Var}\left(E\left(Y_{1,n_{1}+1}\right)-\hat{Y}_{1,n_{1}+1}\right)
&=&
\frac{1}{n_{1}}\lambda_{1}\\
\nonumber\\
\mbox{Var}\left(Y_{1,n_{1}+1}-Y_{1,n_{1}+1}^{\dag}\right)
&=&
\frac{1}{n^2}[(n^{2}+n_{1})\lambda_{1}+n_{2}\lambda_{2}+\ldots+n_{k}\lambda_{k}] \\
\mbox{Var}\left(E\left(Y_{1,n_{1}+1}\right)-Y_{1,n_{1}+1}^{\dag}\right)
&=&
\frac{1}{n^2}[n_{1}\lambda_{1}+n_{2}\lambda_{2}+\ldots+n_{k}\lambda_{k}]
\nonumber\\
\end{eqnarray}

and

\begin{eqnarray}
\nonumber\\
\mbox{MSE}_{1}\left(\hat{Y}_{1,n_{1}+1}\right)
&=&
\left(1+\frac{1}{n_{1}}\right)\lambda_{1} \\
\mbox{MSE}_{2}\left(\hat{Y}_{1,n_{1}+1}\right)
&=&
\frac{1}{n_{1}}\lambda_{1} \\
\nonumber\\
\nonumber\\
\mbox{MSE}_{1}\left(Y_{1,n_{1}+1}^{\dag}\right) &=&
\frac{1}{n^{2}}[(n^{2}-n_{1})\lambda_{1}-n_{2}\lambda_{2}-\ldots-n_{k}\lambda_{k}]^{2} \\
&&+ \frac{1}{n^2} [ (n^{2}+n_{1})\lambda_1 +n_{2}\lambda_{2}
+\ldots+n_{k}\lambda_{k} ]
\nonumber\\
\nonumber\\
\mbox{MSE}_{2}\left(Y_{1,n_{1}+1}^{\dag}\right) &=&
\frac{1}{n^{2}}[n_{1}\lambda_{1}-n_{2}\lambda_{2}-\ldots-n_{k}\lambda_{k}]^{2} \\
&&+ \frac{1}{n^2} [n_{1}\lambda_1 +n_{2}\lambda_{2}
+\ldots+n_{k}\lambda_{k} ]
\end{eqnarray}

One can ask the question: if we want to minimise the MSE, is it better
to use the short baseline or the long baseline?
The answer to this question can be obtained by the following comparison:
\begin{eqnarray}
\mbox{MSE}(Y_{1,n_{1}+1}^{\dag})&<&\mbox{MSE}(\hat{Y}_{1,n_{1}+1})  \\
\end{eqnarray}

Neither side can be evaluated exactly in practice since they both
depend on the unknown parameters $\lambda_1,...,\lambda_k$.
But as a practical approach one could use plug-in estimators for the $\lambda$'s.

\subsection{The mixed baseline model}

We now move on to consider the `mixed baseline model', in which
our prediction consists of a weighted average of the mean hurricane numbers in the $k$ periods,
which we write as:

\begin{equation}
Y_{1,n_{1}+1}^{*}(\alpha)= \alpha_1\hat{\lambda}_{1} +
\alpha_{2}\hat{\lambda}_{2} + \ldots + \alpha_{k}\hat{\lambda}_{k}
\end{equation}

subject to the constraint that the weights add up to 1, i.e.
\begin{equation}
\alpha_{1}+\alpha_{2}+\ldots+\alpha_{k}=1
\end{equation}

We now rewrite the prediction in terms of the real unknown levels of activity, plus noise:

\begin{eqnarray}
Y_{1,n_{1}+1}^{*}(\alpha) &=&
\alpha_{1}\left(\lambda_{1}+\frac{1}{n_{1}}\sum_{j=1}^{n_{1}}\varepsilon_{1,j}\right)
+
\alpha_{2}\left(\lambda_{2}+\frac{1}{n_{2}}\sum_{j=1}^{n_{2}}\varepsilon_{2,j}\right)
+ \ldots +
\alpha_{k}\left(\lambda_{k}+\frac{1}{n_{k}}\sum_{j=1}^{n_{k}}\varepsilon_{k,j}\right)
\nonumber
\end{eqnarray}.

The prediction error is then

\begin{eqnarray}
Y_{1,n_{1}+1}-Y_{1,n_{1}+1}^{*}(\alpha) &=&
(1-\alpha_{1})\lambda_{1}-\alpha_{2}\lambda_{2}-\ldots-\alpha_{3}\lambda_{3} \\
&&+ \varepsilon_{1,n_{1}+1}
-\left(\frac{\alpha_{1}}{n_{1}}\sum_{j=1}^{n_{1}}\varepsilon_{1,j}
+\frac{\alpha_{2}}{n_{2}}\sum_{j=1}^{n_{2}}\varepsilon_{2,j} +\ldots
+\frac{\alpha_{k}}{n_{k}}\sum_{j=1}^{n_{k}}\varepsilon_{k,j}\right)
\end{eqnarray}.

The statistical properties of this prediction are:

\begin{eqnarray}
E\left(Y_{1,n_{1}+1}-Y_{1,n_{1}+1}^{*}(\alpha)\right)
&=&
(1-\alpha_{1})\lambda_{1}-\alpha_{2}\lambda_{2}-\ldots-\alpha_{k}\lambda_{k} \\
\nonumber\\
E\left(E\left(Y_{1,n_{1}+1}\right)-Y_{1,n_{1}+1}^{*}(\alpha)\right)
&=&
(1-\alpha_{1})\lambda_{1}-\alpha_{2}\lambda_{2}-\ldots-\alpha_{k}\lambda_{k}
\end{eqnarray}.

\begin{eqnarray}
\mbox{Var}\left(Y_{1,n_{1}+1}-Y_{1,n_{1}+1}^{*}(\alpha)\right)
&=&
\left(1+\frac{\alpha_{1}^{2}}{n_{1}}\right)\lambda_{1}
+\frac{\alpha_{2}^{2}}{n_{2}}\lambda_{2} +\ldots
+\frac{\alpha_{k}^{2}}{n_{k}}\lambda_{k} \\
\nonumber\\
\mbox{Var}\left(E\left(Y_{1,n_{1}+1}\right)-Y_{1,n_{1}+1}^{*}(\alpha)\right)
&=&
\frac{\alpha_{1}^{2}}{n_{1}}\lambda_{1}
+\frac{\alpha_{2}^{2}}{n_{2}}\lambda_{2} +\ldots
+\frac{\alpha_{k}^{2}}{n_{k}}\lambda_{k}
\end{eqnarray}

and

\begin{eqnarray}
\mbox{MSE}_{1}\left(Y_{1,n_{1}+1}^{*}\right) &=& [(1-\alpha_{1})\lambda_{1}
-\alpha_{2}\lambda_{2} -\ldots
-\alpha_{k}\lambda_{k}]^{2} \\
\nonumber\\
&&+
\left[ \left(1+\frac{\alpha_{1}^{2}}{n_{1}}\right)\lambda_{1}
+\frac{\alpha_{2}^{2}}{n_{2}}\lambda_{2} +\ldots
+\frac{\alpha_{k}^{2}}{n_{k}}\lambda_{k}\right] \\
\nonumber\\
\nonumber\\
\mbox{MSE}_{2}\left(Y_{1,n_{1}+1}^{*}\right) &=& [(1-\alpha_{1})\lambda_{1}
-\alpha_{2}\lambda_{2} -\ldots
-\alpha_{k}\lambda_{k}]^{2} \label{rmse2}\\
\nonumber\\
&&+
\left[\frac{\alpha_{1}^{2}}{n_{1}}\lambda_{1}
+\frac{\alpha_{2}^{2}}{n_{2}}\lambda_{2} +\ldots
+\frac{\alpha_{k}^{2}}{n_{k}}\lambda_{k}\right]
\end{eqnarray}

Our goal is to find the values of the $\alpha$'s that minimize this
MSE subject to

\begin{eqnarray}
\alpha_{1}+\alpha_{2}+\ldots+\alpha_{k}=1 \\
\alpha_{i}\geq0 \\
\mbox{for } i=(1,2,\ldots,k) \mbox{.} \nonumber
\end{eqnarray}

\subsection{Note on definition of the model}\label{annualweights}

We have defined our model in terms of weights on averages over periods
of years. We could also have defined it in terms of weights on
individual years, with these annual weights being constant in each
of the $k$ periods. Since:
\begin{eqnarray}
Y_{1,n_{1}+1}^{*}(\alpha)
&=& \alpha_1\hat{\lambda}_{1} +
\alpha_{2}\hat{\lambda}_{2} + \ldots + \alpha_{k}\hat{\lambda}_{k}\\
&=& \sum_{i=1}^{k} \alpha_i \hat{\lambda}_{i}\\
&=& \sum_{i=1}^{k} \alpha_i \left( \frac{1}{n_{i}}\sum_{j=1}^{n_{i}}Y_{i,j}\right) \\
&=& \sum_{i=1}^{k} \left( \sum_{j=1}^{n_{i}} \frac{\alpha_i
}{n_{i}}Y_{i,j}\right)
\end{eqnarray}

We see that the annual weight on year $j$ in period $i$ is given by
$\frac{\alpha_i}{n_i}$.

These annual weights give us a useful alternative interpretation of the weights
that come out of the model when applied to real data, as we will see below.

\section{Calculating the optimal weights}\label{derive}

For the model described above in section~\ref{four}, how should we
estimate the optimal weights? Ideally we would derive exact expressions
for the weights, that extend the expressions given for the $k=2$ case
in~\citet{j90}. A derivation for exact expressions for the weights, taking into
account the constraint that the weights should sum to one, but
not the constraint that the weights should be positive, is given in appendix 1.
However, when we tested the resulting expressions on real data we found that they gave
negative weights, which are not acceptable.

As a simple alternative for the $k=4$ case we are interested in, we then
resorted to a brute-force
search through all possibilities. If we assume that we only need
accuracy to two decimal places then each weight can only have values
from 0.00 to 1.00, which is 101 possibilities. The fourth weight is
given by the other three, and there are thus only $101^3=1,030,301$
parameter sets to choose from. We have tested this method using
Fortran on a desktop computer, and completing the search for the
weights that give the lowest RMSE took only a few seconds.

There are other (more efficient) numerical methods that one might consider,
especially if $k$ were large, but brute-force searching turned out to
be perfectly adequate for our needs in this particular case.

\section{Results for all hurricanes}\label{results_all}

We now show some results for a number of models applied to the data for all hurricanes,
shown in table~\ref{t01}. Values of bias, standard deviation of errors and RMSE
are all calculated using the four level model, to make them consistent. The RMSE's are shown
as what we call `RMSE2', which is the RMSE on the prediction of the rate rather than the
number of hurricanes (aka the standard error on the rate).
The weights are shown in chronological order, and are shown both
as the original weights used in the model and the annual weights discussed above in section~\ref{annualweights}.
The annual weights are multiplied by 106 to make them easier to understand.

\subsubsection{Model 1: Long term baseline}
The first model we consider is the `long term baseline', which estimates the future rates
as an average of hurricane numbers during the period 1900-2005.
This model gives a prediction of 1.72 hurricanes per year.
This prediction has a low standard deviation (of 0.13), because it is based on many years of data,
but has a significant bias (of 0.56). The bias arises because of the assumptions in our four period model, which are
that the future will come from the same climate as the last 11 years, and the last 11 years have
seen more hurricanes than the 106 years baseline.
The large bias leads to a large RMSE.

\subsubsection{Model 2: Short term baseline}

The second simple model we consider is the `short term baseline', which estimates the future
rates as an average of hurricane numbers during the period 1995-2005.
This model gives a much higher prediction of future hurricane rates, of 2.27 hurricanes per year.
Given our assumptions, this model has no bias. But it does have a large standard deviation of
errors, because of the short period of data used. The resulting RMSE is lower than for the
long term baseline model, but is still rather large (0.45).

The subsequent models are all an attempt to give a better prediction (i.e. a prediction with lower RMSE)
than either of these two very simple models.

\subsubsection{Model 3: Optimal combination of the long and short baselines}

Our next model considers the optimal combination of the long and
short baselines, just as in \citet{j90}, but now with the weights
calculated to minimise the RMSE as given by equation~\ref{rmse2} above (i.e.
to minimise the RMSE calculated in the context of the four level
model). The optimal weights come out to be $66\%$ on the recent
data and $34\%$ on the earlier data. The annual weights show this
to be equivalent to putting just over 16 times as much weight on
the values from the recent years than the earlier years. The RMSE
in this model cannot be worse than the two models that go into it,
by definition. In fact, it is somewhat better than the short
baseline model.

One note: it is important to realise that this model does not assume that there is a $66\%$
chance that the next year will be at the same level as the last 11 years,
and a $34\%$ chance that the next year will revert to the long term baseline.
It assumes that there is a $100\%$ chance that the next year will be at the same
level as the last 11 years.
The weighting comes in as a mathematical result of trying to reduce the errors in our prediction.
If one wanted to include some probability of reverting to a lower level in the future,
one would need to do that some other way.

\subsubsection{Model 4: Average of the active periods}

As mentioned in the introduction, no-one really believes that the period 1900-1994 is uniform and without fluctuations
in the hurricane rate. Model 4 is the first to try and take that into account. It does so by ignoring
the two periods with lower rates, and taking a straight average of the two periods with a higher rate.
This model performs well in terms of RMSE, and is the best model so far.

\subsubsection{Model 5: Optimal combination of the active periods}

However, the earlier active period was not as active as the recent active period, and so perhaps we should
put more weight on the recent period than comes from the straight average of all active years that we used in model 4.
Model 5 tries an optimal combination of the two active periods to capture this.
We get a lower RMSE than model 4, as we must by construction (because we've added an extra parameter)
and slightly higher weights on the recent data.
The forecast is slightly higher than model 4 because of these higher weights on the recent
(and more active) period.

\subsubsection{Model 6: One active period, one inactive period}

A potential shortcoming of models 4 and 5 is that they don't use the whole available data set.
We now attempt ways to use the whole data set, but still resolve the idea that there was an
active period in the middle of the last century. Model 6 does this in the simplest way possible
by aggregating the two less active phases into one period, and the two active phases into another period.
It then combines the average levels across these two aggregated periods with weights that minimise RMSE.
The resulting weights are very interesting:
the algorithm that determines the optimal weights actually sets the weights on the aggregated inactive
period to zero, and model 6 thus becomes equivalent to a straight average of the two active periods.
This is telling us that there is \emph{not} a potential benefit to be had from using the data from the inactive
periods: although using such data would reduce the contribution to the RMSE from the standard deviation of the errors,
it would increase the contribution to the RMSE from the bias by more.

\subsubsection{Model 7: Two active periods, one inactive period}

An extension of model 6 above is to distinguish between the rates in the two active periods, but
not in the two inactive periods. As might be expected given the results for model 6, this method
again drops the inactive data from the combination, and thus reverts to model 5.

\subsubsection{Model 8: four distinct periods}

The most complex model we consider takes four periods at different levels and combines them
using four independent weights (subject to the constraints). By definition this model must
give the best RMSE score, and it does. However, we suspect that the score may not be
not \emph{statistically significantly} lower
than the scores from models 5 and 7, and the forecast is very similar. Model 8
gives a zero weight to the data from period 2, and only a very small weight to the data from
the earlier inactive period.

\subsection{Discussion}

Consideration of these models has yielded the interesting result that the data from the two
inactive periods cannot be used in a very useful way to improve our forecast, because the
rates during the inactive periods are too different. The data from the earlier active period,
however, can definitely be used. Our best model in terms of RMSE (model 8) does put a small weight on
the data from the earlier inactive period. However, the RMSE benefit is very small, and this model
has 3 parameters versus only 1 parameter for model 5. If we did an analysis of whether the
benefit from using model 8 over model 5 were statistically significant, we suspect it wouldn't be.

\section{Results for intense hurricanes}\label{results_intense}

The results for intense hurricanes (shown in table~\ref{t02}) are broadly similar, but with some specific differences.
We only comment on the differences.

\subsubsection{Model 2: short term baseline}

It is interesting to note that the short term baseline
model now does worse than the long term baseline model, which is a reverse of the results for total
hurricane numbers.
This is because there are fewer cat 3-5 storms,
and using only 11 years of data works less well than it does for cat 1-5 storms.

\subsubsection{Model 5: optimal combination of the two active periods}

The optimal combination of the two active periods is now more or less the same as the straight average.
This is because the optimal combination puts more weight on the earlier period than it does for cat 1-5 hurricanes.
This is again because there are fewer cat 3-5 storms,
so we need to use more years of data to estimate them accurately, even if that data is not ideal.

\section{Results for basin hurricanes}\label{results_basin}

For future reference, we also include results for basin hurricanes, in tables~\ref{t03} and~\ref{t04}.

%

\section{Conclusions}\label{conclusions}

We have considered the question of how to combine historical hurricane numbers to make a prediction
of the future. We have made this question mathematically tractable by assuming that the hurricane
rates of the last 106 years were characterized by four periods of differing levels of activity, and that
future activity will continue at the current level.
We then look at different ways that one might combine estimates of activity from these four periods.

There are various conclusions we come to.
Firstly, using data from the two inactive periods doesn't add much, if anything, to our predictions. This is
because the levels of activity during the inactive periods were low relative to the current level
of activity. Any potential benefit from using more data is wiped out by the fact that the data is
at the wrong level. Secondly, using data from the earlier inactive period can definitely help
our predictions. This data is not at exactly the same level as the current period of high activity,
but is sufficiently close that the bias introduced by using this data is small compared with the
beneficial reduction in variance. Overall, using this data reduces the RMSE of our predictions
considerably.

The biggest flaw in this study is that we don't have an objective method for choosing the best
model from the 8 models we have considered. The 3 parameter model (model 8) gives the best
results (by construction), but may be overfitted relative to simpler models such as model 4 or model 5.
Our next goal is therefore to consider how to select between these models.


\appendix
\section{Exact derivation for weights for arbitrary $k$}

We derive exact expressions for the weights as follows. We define the
problem as needing to minimise a cost function $L$, given below.
This cost function includes a Lagrange multiplier to account for the
constraint that the weights must sum to 1.

\begin{equation}
L=
\left[\left[(1-\alpha_{1})\lambda_{1}
-\alpha_{2}\lambda_{2} -\ldots
-\alpha_{k}\lambda_{k}\right]^{2}
+ \left[\left(1+\frac{\alpha_{1}^{2}}{n_{1}}\right)\lambda_{1}
+\frac{\alpha_{2}^{2}}{n_{2}}\lambda_{2} +\ldots
+\frac{\alpha_{k}^{2}}{n_{k}}\lambda_{k}\right]
- \gamma\left(\sum_{i=1}^{k}\alpha_{i}-1\right)\right]
\end{equation}

We then differentiate this cost function
by the weight vector
$\alpha$ and the lagrange multiplier $\gamma$, giving a $k+1$ by 1 matrix of derivatives:

\begin{equation}
\left(
  \begin{array}{c}
    \frac{\partial}{\partial\alpha}\mbox{MSE}[\hat{Y}_{1,n_{1}+1}^{*}] \\
    \\
    \frac{\partial}{\partial\gamma}\mbox{MSE}[\hat{Y}_{1,n_{1}+1}^{*}] \\
  \end{array}
\right)
=
\left(
  \begin{array}{c}
    -2\lambda_{1}[(1-\alpha_{1})\lambda_{1}-\lambda_{2}\alpha_{2}-\ldots-\lambda_{k}\alpha_{k}]+2\lambda_{1}\frac{\alpha_{1}}{n_{1}}-\gamma \\
    \\
    -2\lambda_{2}[(1-\alpha_{1})\lambda_{1}-\lambda_{2}\alpha_{2}-\ldots-\lambda_{k}\alpha_{k}]+2\lambda_{1}\frac{\alpha_{2}}{n_{2}}-\gamma \\
    \\
    \vdots \\
    \\
    -2\lambda_{k}[(1-\alpha_{1})\lambda_{1}-\lambda_{2}\alpha_{2}-\ldots-\lambda_{k}\alpha_{k}]+2\lambda_{1}\frac{\alpha_{k}}{n_{k}}-\gamma \\
    \\
    -\alpha_{1}-\alpha_{2}-\ldots-\alpha_{k}+1
  \end{array}
\right)
\end{equation}

To find the minimum, we set:

\begin{eqnarray}
\left(
  \begin{array}{c}
    \frac{\partial}{\partial\alpha}\mbox{MSE}[\hat{Y}_{1,n_{1}+1}^{*}] \\
    \\
    \frac{\partial}{\partial\gamma}\mbox{MSE}[\hat{Y}_{1,n_{1}+1}^{*}] \\
  \end{array}
\right) = 0
\end{eqnarray}

and rearranging the equations we obtain

\begin{eqnarray}
    -2\lambda_{1}\lambda_{1} + 2\left(\lambda_{1}\lambda_{1}+\lambda_{1}\frac{1}{n_{1}}\right)\alpha_{1} + 2\lambda_{1}\lambda_{2}\alpha_{2} + \ldots + 2\lambda_{1}\lambda_{k}\alpha_{k} - \gamma &=& 0 \\
    -2\lambda_{1}\lambda_{2} + 2\lambda_{1}\lambda_{2}\alpha_{1} + 2\left(\lambda_{2}\lambda_{2}+\lambda_{1}\frac{1}{n_{2}}\right)\alpha_{2} + \ldots + 2\lambda_{2}\lambda_{k}\alpha_{k} - \gamma &=& 0 \\
    &\vdots& \\
    -2\lambda_{1}\lambda_{k} + 2\lambda_{1}\lambda_{k}\alpha_{1} + 2\lambda_{k}\lambda_{2}\alpha_{2} + \ldots + 2\left(\lambda_{k}\lambda_{k}+\lambda_{1}\frac{1}{n_{k}}\right)\alpha_{k} - \gamma &=& 0 \\
    \nonumber \\
    1 - \alpha_{1} - \alpha_{2} - \ldots - \alpha_{k} - 0\cdot\gamma &=& 0
\end{eqnarray}

This set of equations can be rewritten in matrix form as

\begin{eqnarray}
    \left(
      \begin{array}{ccccc}
        2\left(\lambda_{1}\lambda_{1}+\lambda_{1}\frac{1}{n_{1}}\right) & 2\lambda_{1}\lambda_{2} & \cdots & 2\lambda_{1}\lambda_{k} & -1 \\
        2\lambda_{1}\lambda_{2} & 2\left(\lambda_{2}\lambda_{2}+\lambda_{1}\frac{1}{n_{2}}\right) & \cdots & 2\lambda_{2}\lambda_{k} & -1 \\
        \vdots & \vdots & \ddots & \vdots & \vdots \\
        2\lambda_{1}\lambda_{k} & 2\lambda_{2}\lambda_{k} & \cdots & 2\left(\lambda_{k}\lambda_{k}+\lambda_{1}\frac{1}{n_{k}}\right) & -1 \\
        \\
        -1 & -1 & \cdots & -1 & 0 \\
      \end{array}
    \right)
    \times
    \left(
      \begin{array}{c}
        \alpha_{1} \\
        \alpha_{2} \\
        \vdots \\
        \alpha_{k} \\
        \gamma \\
      \end{array}
    \right)
    &=&
    \left(
      \begin{array}{c}
        2\lambda_{1}\lambda_{1} \\
        2\lambda_{1}\lambda_{2} \\
        \vdots \\
        2\lambda_{1}\lambda_{k} \\
        -1 \\
      \end{array}
    \right)
\end{eqnarray}

In shorthand notation we can write

\begin{equation}
M\times \alpha=c
\end{equation}

where $M$ is a symmetric $(k+1)\times(k+1)$ matrix of coefficients, $\alpha$ is $k+1$ column vector of weights $\alpha_{i}$ and Lagrange multiplier $\gamma$ and $c$ is a $k+1$ column vector of constants.

The solution for alpha is then obtained by

\begin{equation}
    \alpha=M^{-1}\times c
\end{equation}

\subsection{Special cases}

The simplest interesting case is when $k=2$, which is equivalent to the original model of~\citet{j90}.
We now show that the above expressions agree with that model for this special case.

Setting $k=2$, we get

\begin{eqnarray}
\alpha &=&
\left(
  \begin{array}{c}
    \alpha_{1} \\
    \alpha_{2} \\
    \gamma \\
  \end{array}
\right) \\
\nonumber \\
M &=&
\left(
  \begin{array}{ccc}
    2\left(\lambda_{1}\lambda_{1}+\lambda_{1}\frac{1}{n_{1}}\right) & 2\lambda_{1}\lambda_{2} & -1 \\
    2\lambda_{2}\lambda_{1} & 2\left(\lambda_{2}\lambda_{2}+\lambda_{1}\frac{1}{n_{2}}\right) & -1 \\
    -1 & -1 & 0 \\
  \end{array}
\right) \\
\nonumber \\
c &=&
\left(
  \begin{array}{c}
    2\lambda_{1}\lambda_{1} \\
    2\lambda_{1}\lambda_{2} \\
    -1 \\
  \end{array}
\right)
\end{eqnarray}

and so

\begin{equation}
M^{-1}
=
\left(
  \begin{array}{ccc}
    \frac{1}{2\left[\frac{\lambda_{1}}{n_{1}}+\left(\lambda_{1}-\lambda_{2}\right)^{2}\right]} & -\frac{1}{2\left[\frac{\lambda_{1}}{n_{1}}+\left(\lambda_{1}-\lambda_{2}\right)^{2}\right]} & -\frac{\frac{\lambda_{2}}{n_{2}}-\lambda_{1}\lambda_{2}+\lambda_{2}^{2}}{\frac{\lambda_{1}}{n_{1}}+\left(\lambda_{1}-\lambda_{2}\right)^{2}} \\
    -\frac{1}{2\left[\frac{\lambda_{1}}{n_{1}}+\left(\lambda_{1}-\lambda_{2}\right)^{2}\right]} & \frac{1}{2\left[\frac{\lambda_{1}}{n_{1}}+\left(\lambda_{1}-\lambda_{2}\right)^{2}\right]} & -\frac{\frac{\lambda_{1}}{n_{1}}-\lambda_{1}+\lambda_{1}^{2}}{\frac{\lambda_{1}}{n_{1}}+\left(\lambda_{1}-\lambda_{2}\right)^{2}} \\
    -\frac{\frac{\lambda_{2}}{n_{2}}-\lambda_{1}\lambda_{2}+\lambda_{2}^{2}}{\frac{\lambda_{1}}{n_{1}}+\left(\lambda_{1}-\lambda_{2}\right)^{2}} & -\frac{\frac{\lambda_{1}}{n_{1}}-\lambda_{1}\lambda_{2}+\lambda_{1}^{2}}{\frac{\lambda_{1}}{n_{1}}+\left(\lambda_{1}-\lambda_{2}\right)^{2}} & -2\frac{\frac{\lambda_{1}\lambda_{2}}{n_{1}n_{2}}+\frac{\lambda_{1}^2\lambda_{2}}{n_{2}}+\frac{\lambda_{1}\lambda_{2}^{2}}{n_{1}}}{\frac{\lambda_{1}}{n_{1}}+\left(\lambda_{1}-\lambda_{2}\right)^{2}} \\
  \end{array}
\right)
\end{equation}

and

\begin{equation}
\alpha =
\frac{1}{n_{1}n_{2}\left(\lambda_{1}-\lambda_{2}\right)^2+n_{2}\lambda_{1}+n_{1}\lambda_{2}}
\left(
  \begin{array}{c}
    n_{1}n_{2}\left(\lambda_{1}-\lambda_{2}\right)^{2}+n_{1}\lambda_{2} \\
    \\
    n_{2}\lambda_{1} \\
    \\
    2\left[\lambda_{1}\lambda_{2}+n_{2}\left(\lambda_{1}\lambda_{2}^{2}-\lambda_{1}^{2}\lambda_{2}\right)\right] \\
  \end{array}
\right)
\end{equation}

The first two weights in this expression agree with the corresponding expression in~\citet{j90}.

\bibliography{arxiv}

\newpage
 \setlength{\hoffset}{-1.6in}\setlength{\oddsidemargin} {2.0cm}
 \setlength{\textwidth}{12cm} \setlength{\voffset}  {-1in}
 \setlength{\topmargin}{1cm} \setlength              {\textheight}{25cm}
 \setlength{\unitlength}{1cm} \setlength             {\parindent}{0cm}

 \begin{table}[h!]
   \centering
 \begin{tabular}{|c|c|c|c|c|c|c|c|}
  \hline
  1 & 2 & 3 & 4 & 5 & 6 & 7 & 8\\
  \hline
 Model & Model Name(No. Yrs)& Fcst & RMSE2               & Pctage & Bias,SD & Weights (a4,a3,a2,a1) & Scaled Annual      Weights \\
  \hline
        1& long-term BL (106)       & 1.72&0.570&  33.21&  0.56, 0.13& 0.41, 0.21, 0.28, 0.10& 1.00, 1.00, 1.00, 1.00  \\
         2& short-term BL (11)      & 2.27&0.455&  20.00&  0.00, 0.45& 0.00, 0.00, 0.00, 1.00& 0.00, 0.00, 0.00, 9.64  \\
    3& 2 pds, pre/post 1994/5 (106) & 2.06&0.369&  17.92&  0.21, 0.30& 0.15, 0.08, 0.11, 0.66& 0.38, 0.38, 0.38, 6.36  \\
               4&2 active pds  (33) & 2.06&0.328&  15.91&  0.21, 0.25& 0.00, 0.67, 0.00, 0.33& 0.00, 3.21, 0.00, 3.21  \\
  5& 2 active pds, optimal (33) & 2.11&0.315&  14.93&  0.17, 0.27& 0.00, 0.52, 0.00, 0.48& 0.00, 2.51, 0.00, 4.62  \\
   6& 2 periods,1-Act/1-Inact (106) & 2.06&0.328&  15.91&  0.21, 0.25& 0.00, 0.67, 0.00, 0.33& 0.00, 3.21, 0.00, 3.21  \\
       7& 3 pds,2-Act/1-Inact (106) & 2.11&0.315&  14.93&  0.17, 0.27& 0.00, 0.52, 0.00, 0.48& 0.00, 2.51, 0.00, 4.62  \\
                     8& 4 pds (106) & 2.09&0.312&  14.97&  0.19, 0.25& 0.09, 0.43, 0.00, 0.47& 0.23, 2.09, 0.00, 4.55  \\
 \hline
 \end{tabular}
 \caption{
Mixed baseline results for cat 1-5 US landfalling hurricanes
 }\label{t01}
 \end{table}

 \begin{table}[h!]
   \centering
 \begin{tabular}{|c|c|c|c|c|c|c|c|}
  \hline
  1 & 2 & 3 & 4 & 5 & 6 & 7 & 8\\
  \hline
 Model & Model Name(No. Yrs)& Fcst & RMSE2               & Pctage & Bias,SD & Weights (a4,a3,a2,a1) & Scaled Annual      Weights \\
  \hline
        1& long-term BL (106)       & 0.65&0.270&  41.44&  0.26, 0.08& 0.41, 0.21, 0.28, 0.10& 1.00, 1.00, 1.00, 1.00  \\
         2& short-term BL (11)      & 0.91&0.287&  31.62&  0.00, 0.29& 0.00, 0.00, 0.00, 1.00& 0.00, 0.00, 0.00, 9.64  \\
    3& 2 pds, pre/post 1994/5 (106) & 0.77&0.207&  26.89&  0.14, 0.15& 0.22, 0.11, 0.15, 0.52& 0.54, 0.54, 0.54, 5.01  \\
               4&2 active pds  (33) & 0.85&0.171&  20.20&  0.06, 0.16& 0.00, 0.67, 0.00, 0.33& 0.00, 3.21, 0.00, 3.21  \\
  5& 2 active pds, optimal (33) & 0.85&0.171&  20.14&  0.06, 0.16& 0.00, 0.64, 0.00, 0.36& 0.00, 3.11, 0.00, 3.42  \\
   6& 2 periods,1-Act/1-Inact (106) & 0.83&0.170&  20.49&  0.08, 0.15& 0.04, 0.62, 0.03, 0.31& 0.10, 2.98, 0.10, 2.98  \\
       7& 3 pds,2-Act/1-Inact (106) & 0.83&0.169&  20.41&  0.08, 0.15& 0.05, 0.57, 0.03, 0.35& 0.12, 2.76, 0.12, 3.33  \\
                     8& 4 pds (106) & 0.82&0.165&  20.19&  0.09, 0.14& 0.15, 0.52, 0.00, 0.33& 0.37, 2.49, 0.00, 3.20  \\
 \hline
 \end{tabular}
 \caption{
Mixed baseline results for cat 3-5 US landfalling hurricanes
 }\label{t02}
 \end{table}

 \begin{table}[h!]
   \centering
 \begin{tabular}{|c|c|c|c|c|c|c|c|}
  \hline
  1 & 2 & 3 & 4 & 5 & 6 & 7 & 8\\
  \hline
 Model & Model Name(No. Yrs)& Fcst & RMSE2               & Pctage & Bias,SD & Weights (a4,a3,a2,a1) & Scaled Annual      Weights \\
  \hline
        1& long-term BL (106)       & 5.28&3.179&  60.18&  3.17, 0.22& 0.41, 0.21, 0.28, 0.10& 1.00, 1.00, 1.00, 1.00  \\
         2& short-term BL (11)      & 8.45&0.877&  10.37&  0.00, 0.88& 0.00, 0.00, 0.00, 1.00& 0.00, 0.00, 0.00, 9.64  \\
    3& 2 pds, pre/post 1994/5 (106) & 8.25&0.851&  10.32&  0.21, 0.83& 0.03, 0.01, 0.02, 0.94& 0.06, 0.06, 0.06, 9.08  \\
               4&2 active pds  (33) & 6.91&1.612&  23.33&  1.55, 0.46& 0.00, 0.67, 0.00, 0.33& 0.00, 3.21, 0.00, 3.21  \\
  5& 2 active pds, optimal (33) & 8.18&0.823&  10.06&  0.28, 0.77& 0.00, 0.12, 0.00, 0.88& 0.00, 0.58, 0.00, 8.48  \\
   6& 2 periods,1-Act/1-Inact (106) & 6.91&1.612&  23.33&  1.55, 0.46& 0.00, 0.67, 0.00, 0.33& 0.00, 3.21, 0.00, 3.21  \\
       7& 3 pds,2-Act/1-Inact (106) & 8.18&0.823&  10.06&  0.28, 0.77& 0.00, 0.12, 0.00, 0.88& 0.00, 0.58, 0.00, 8.48  \\
                     8& 4 pds (106) & 8.18&0.823&  10.06&  0.28, 0.77& 0.00, 0.12, 0.00, 0.88& 0.00, 0.58, 0.00, 8.48  \\
 \hline
 \end{tabular}
 \caption{
Mixed baseline results for cat 1-5 Atlantic basin hurricanes
 }\label{t03}
 \end{table}

 \begin{table}[h!]
   \centering
 \begin{tabular}{|c|c|c|c|c|c|c|c|}
  \hline
  1 & 2 & 3 & 4 & 5 & 6 & 7 & 8\\
  \hline
 Model & Model Name(No. Yrs)& Fcst & RMSE2               & Pctage & Bias,SD & Weights (a4,a3,a2,a1) & Scaled Annual      Weights \\
  \hline
        1& long-term BL (106)       & 2.17&1.926&  88.78&  1.92, 0.14& 0.41, 0.21, 0.28, 0.10& 1.00, 1.00, 1.00, 1.00  \\
         2& short-term BL (11)      & 4.09&0.610&  14.91&  0.00, 0.61& 0.00, 0.00, 0.00, 1.00& 0.00, 0.00, 0.00, 9.64  \\
    3& 2 pds, pre/post 1994/5 (106) & 3.93&0.587&  14.93&  0.16, 0.56& 0.03, 0.02, 0.02, 0.93& 0.08, 0.08, 0.08, 8.91  \\
               4&2 active pds  (33) & 3.64&0.563&  15.48&  0.45, 0.33& 0.00, 0.67, 0.00, 0.33& 0.00, 3.21, 0.00, 3.21  \\
  5& 2 active pds, optimal (33) & 3.84&0.482&  12.57&  0.26, 0.41& 0.00, 0.38, 0.00, 0.63& 0.00, 1.81, 0.00, 6.02  \\
   6& 2 periods,1-Act/1-Inact (106) & 3.64&0.563&  15.48&  0.45, 0.33& 0.00, 0.67, 0.00, 0.33& 0.00, 3.21, 0.00, 3.21  \\
       7& 3 pds,2-Act/1-Inact (106) & 3.84&0.482&  12.57&  0.26, 0.41& 0.00, 0.38, 0.00, 0.63& 0.00, 1.81, 0.00, 6.02  \\
                     8& 4 pds (106) & 3.83&0.482&  12.57&  0.26, 0.41& 0.00, 0.38, 0.00, 0.62& 0.00, 1.81, 0.00, 6.02  \\
 \hline
 \end{tabular}
 \caption{
Mixed baseline results for cat 3-5 Atlantic basin hurricanes
 }\label{t04}
 \end{table}

\end{document}